\documentclass[12pt]{iopart}

\usepackage{iopams}  

\usepackage{graphicx}
\usepackage{xcolor}
\usepackage{hyperref}
\usepackage{siunitx}
\begin{document}
\title{A New Design for a Traveling-Wave Zeeman Decelerator: II. Experiment}

\author{Tomislav Damjanovi\'c$^{1}$, Stefan Willitsch$^{1}$, Nicolas Vanhaecke$^{2,3,\ddagger}$, Henrik Haak$^{2}$, Gerard Meijer$^{2}$, JeanPaul Cromi\'eres$^{3}$, Dongdong Zhang$^{4,1,2,*}$}
 \address{$^{1}$ 
 Department of chemistry, University of Basel, Klingelbergstrasse 80, 4056 Basel, Switzerland}
 \address{$^{2}$
 Fritz-Haber-Institut der Max-Planck-Gesellschaft, Faradayweg 4-6, 14195 Berlin, Germany
}
\address{$^{3}$Labaratorie Aim\'e Cotton, CNRS, Universit\'e Paris-Sud, 91405 Orsay, France}

 \address{$^{4}$Institute of atomic and molecular physics, Jilin university, 2699 Qianjin Avenue, Changchun city, 130012, China}
\ead{$^{*}$dongdongzhang@jlu.edu.cn}
\footnote{Present address: European Patent Office, Patentlaan 2, 2288 EE Rijswijk, The Netherlands}
\vspace{10pt}
\begin{indented}
\item[] \today
\end{indented}
\begin{abstract}
A novel traveling-wave Zeeman decelerator based on a double-helix coil geometry capable of decelerating paramagnetic molecules with high efficiency is presented. Moving magnetic traps are generated by applying time-dependent currents through the decelerator coils. Paramagnetic molecules in low-field-seeking Zeeman states are confined inside the moving traps which are decelerated to lower forward velocities. As a prototypical example, we demonstrate the deceleration of OH radicals from an initial velocity of 445 m/s down to various final velocities. The experimental results are analyzed and numerically reproduced with the help of trajectory simulations confirming the phase-space stability and efficiency of the deceleration of the molecules in the new device.
\\
\end{abstract}

\maketitle

\section{Introduction}

Translationally cold molecules represent ideal systems for precision measurements~\cite{safronova18a,cairncross19a,chupp19a} and fundamental collision studies~\cite{caldwell20,cheuk20,jongh20,segev19,gao18a,vogels18a,onvlee17a,Klein17a,doyle16a}. Over the last two decades, a variety of methods has been developed to produce cold molecules from warm samples, including buffer-gas cooling~\cite{truppe18a}, Stark\cite{bethlem1999a,Haas2017a} and Zeeman deceleration~\cite{vanhaecke07a,Narevicius07a,hogan11a,narevicius12a,meerakker12a}, centrifugal-force based deceleration~\cite{chervenkov14a} and direct laser cooling~\cite{kozyryev17a,truppe17a,coherent18a,lim18a,jarvis18a,tarbutt18a,baum20a,mitra20}. In addition, a number of "indirect" methods have been implemented for "synthesizing" cold molecules from cold atoms including  photoassociation~\cite{kevin06a} and magnetoassociation\cite{chin10a}.

Over the last decade, Zeeman deceleration has been established as one of the most widely used methods for generating samples of translationally cold molecules~\cite{vanhaecke07a,Narevicius07a,Narevicius07b}. Zeeman deceleration relies on the slowing down of beams of paramagnetic atoms or molecules in low-field-seeking (LFS) Zeeman  states which are inserted into a linear array of coils generating time-dependent inhomogeneous magnetic fields~\cite{Lavert_Ofir11a,wiederkehr11a,wiederkehr12a,motsch14a,Akerman15a,plomp19a}. Cold atoms and molecules produced by Zeeman deceleration have found applications in precision measurements~\cite{semeria18a}, fundamental studies of collisions between molecules at very low temperatures~\cite{segev19a} and trapping studies~\cite{narevicius12a,liu15a,liu17a,akerman17a}.
Conventional Zeeman deceleration techniques rely on reducing the kinetic energy of atoms or molecules in LFS states by a fast switching of time-dependant inhomogeneous magnetic fields. On the other hand, traveling-wave Zeeman deceleration relies on traveling inhomogeneous magnetic waves, which features the same working principle as a traveling-wave Stark decelerator~\cite{meek08a,meek09a,meek09b,osterwalder10a,meek11a,bulleid12a} but works for different types of atoms and molecules. Atoms or molecules in their LFS states are trapped around the minima of the traveling magnetic wave which is slowed down to reduce the kinetic energy of the trapped particles.  
Conventional Zeeman decelerators experience problems of the coupling of the longitudinal and transverse motions of the decelerated molecular beam similar to conventional Stark decelerators~\cite{meerakker12a}. These couplings can severely impair the deceleration efficiency~\cite{wiederkehr11a,Zhang2016a,reens20a}. Traveling-wave Zeeman decelerators \cite{Narevicius07a,mcard18a} overcome the losses due to non-ideal transverse focusing forces, as do designs based on electromagnetic coils combined with static magnetic hexapoles~\cite{cremers17a}. 

In this work, we introduce a new traveling-wave Zeeman decelerator based on a double-helix coil geometry capable of decelerating samples of paramagnetic species down to an arbitrary final velocity. Compared to the conventional Zeeman or Stark decelerators, the presented decelerator exhibits full three-dimensional confinement of the molecules at a full range of velocities leading to an improvement of the overall phase-space acceptance. 
In~\Sref{section:A traveling magnetic wave}, we describe the working principle of the decelerator. In~\Sref{section:Implementation}, the physical implementation of the decelerator modules is detailed. The electronics for generating the time-dependent currents giving rise to the traveling-wave magnetic traps is described in~\Sref{section:Electronics}. The principle of the decelerator is proven with experimental results backed up by simulations in~\Sref{section:experiment} followed by conclusions and an outlook.

\section{Generation of traveling magnetic waves}
\label{section:A traveling magnetic wave}

\begin{figure}
\centering
\includegraphics[width=6.6cm]{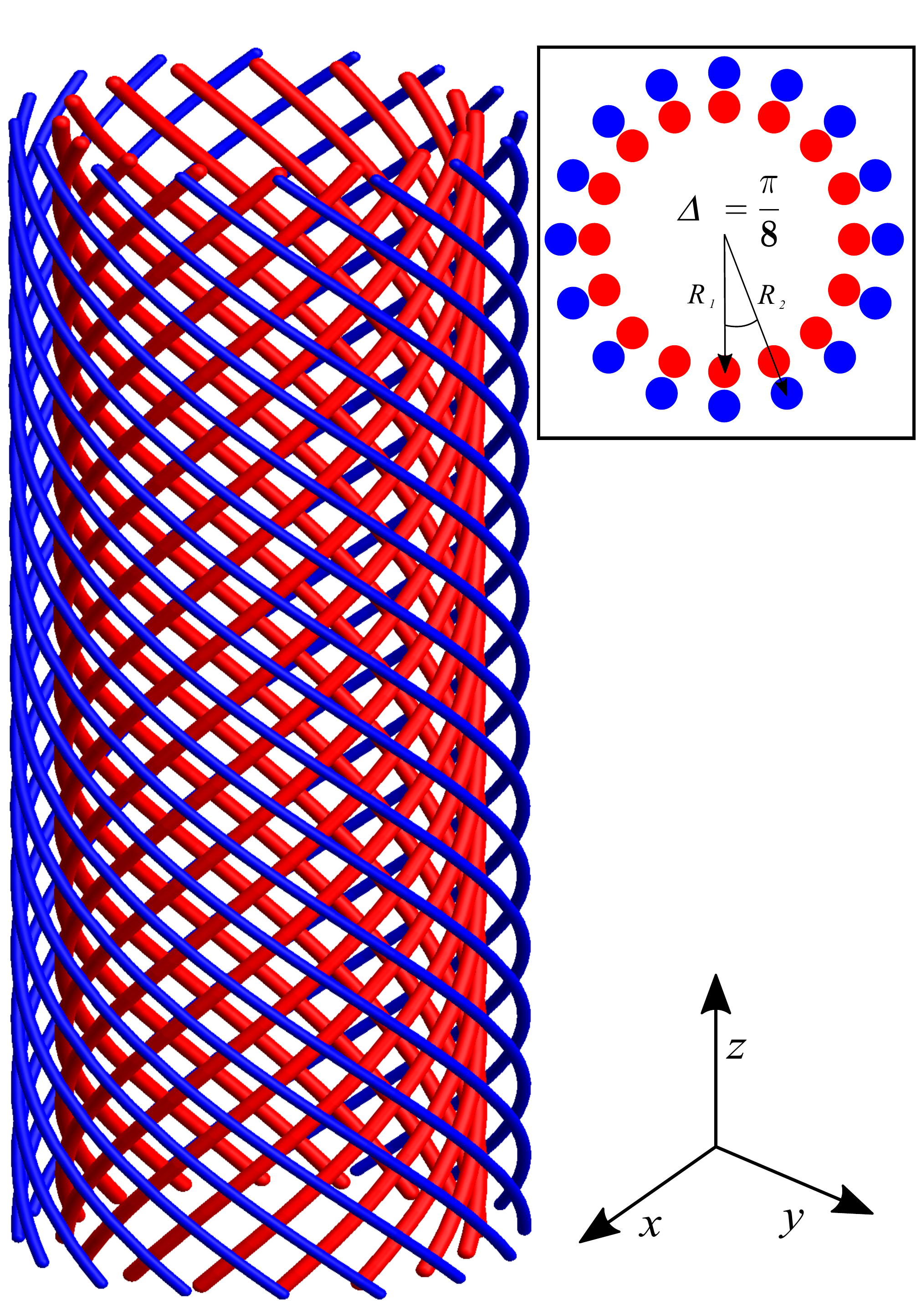}
\caption{Schematic of the double-helix coil geometry used in the present Zeeman decelerator. Two layers of 16 wires each are wound around a cylindrical base in right-handed (red) and left-handed (blue) helices. The thickness of the wires has been scaled down for clarity. The inset shows a schematic of the initial positions of the wires on the front face of a coil module.}
\label{fig:fig1}
\end{figure}

We start by giving a brief summary of the principle for generating traveling magnetic waves in the present device. A full account of the theory is published in an accompanying paper \cite{damjanovic21a}. Let us consider a coil geometry consisting of two layers of wires wound around a cylindrical base (\Fref{fig:fig1}). The first layer consists of 16 wires wound in right-handed (RH) oriented helices superimposed by a second layer of 16 left-handed (LH) oriented helices. The inset in~\Fref{fig:fig1} shows a cross section of the helix assembly. Wires are distributed along a circle of radius $R$, where $R=R_1$ corresponds to the radius of the right-handed helix layer and $R=R_2$ to the left-handed helix layer. Neighbouring helices in each layer are separated by an arc length of $R\Delta$ where $\Delta = \pi/8$. Applying time-dependent currents of the form 

\begin{eqnarray}
&I_{\mathrm{RH,n}}(\mathrm{t}) = \mathit{I}_{\mathrm{RH}}\sin(\phi_{\mathrm{RH}}(t)+\mathrm{n}\Delta)
\\
&I_{\mathrm{LH,n}}(\mathrm{t}) = I_{\mathrm{LH}}\sin(\phi_{\mathrm{LH}}(t)+\mathrm{n}\Delta)
\end{eqnarray}
to the $n$-th wire ($n=0,...,15$) in each layer leads to a total magnetic field $\mathbf{B}$ of the form of a traveling wave on the central axis of the decelerator: 
\begin{eqnarray}
|\mathbf{B}(x=0,y=0,z,t)| = |B_0\sin(kz-\phi_z(t))|.
\end{eqnarray}
Here, $I_{\mathrm{RH}}$ and $I_\mathrm{{LH}}$ 
denote the current in the right and left handed layer, respectively and $\phi_{\mathrm{RH}}(t)$ and $\phi_{\mathrm{LH}}(t)$ stand for the corresponding phases. $\phi_{\mathrm{RH}}(t)$ and $\phi_{\mathrm{LH}}(t)$ are related to $\phi_z(t)$ and $\phi_\theta(t)$ by 

\begin{eqnarray}
&\phi_{\mathrm{RH}}(t) = \phi_z(t)+\phi_\theta(t),
\\
&\phi_{\mathrm{LH}}(t) = \phi_z(t)-\phi_\theta(t),
\end{eqnarray}

where $\phi_z(t)=at^2+bt+c$ defines the time-dependent position of the minimum of the magnetic field, and $\phi_\theta(t)$ controls the orientation of the magnetic field in the $xy$ plane perpendicular to the longitudinal direction $z$ \cite{damjanovic21a}. The parameters $a/k$ and $b/k$ denote deceleration (or acceleration) and initial velocity, respectively, of the traveling wave. Here, $k=2\pi/\lambda$, where $\lambda=14\ $mm is the periodicity of the helices. $c$ is a free parameter, corresponding to an arbitrary phase shift of the traveling wave at time $t = 0$~s. 

\begin{figure}
\centering
\includegraphics[width=6.6cm]{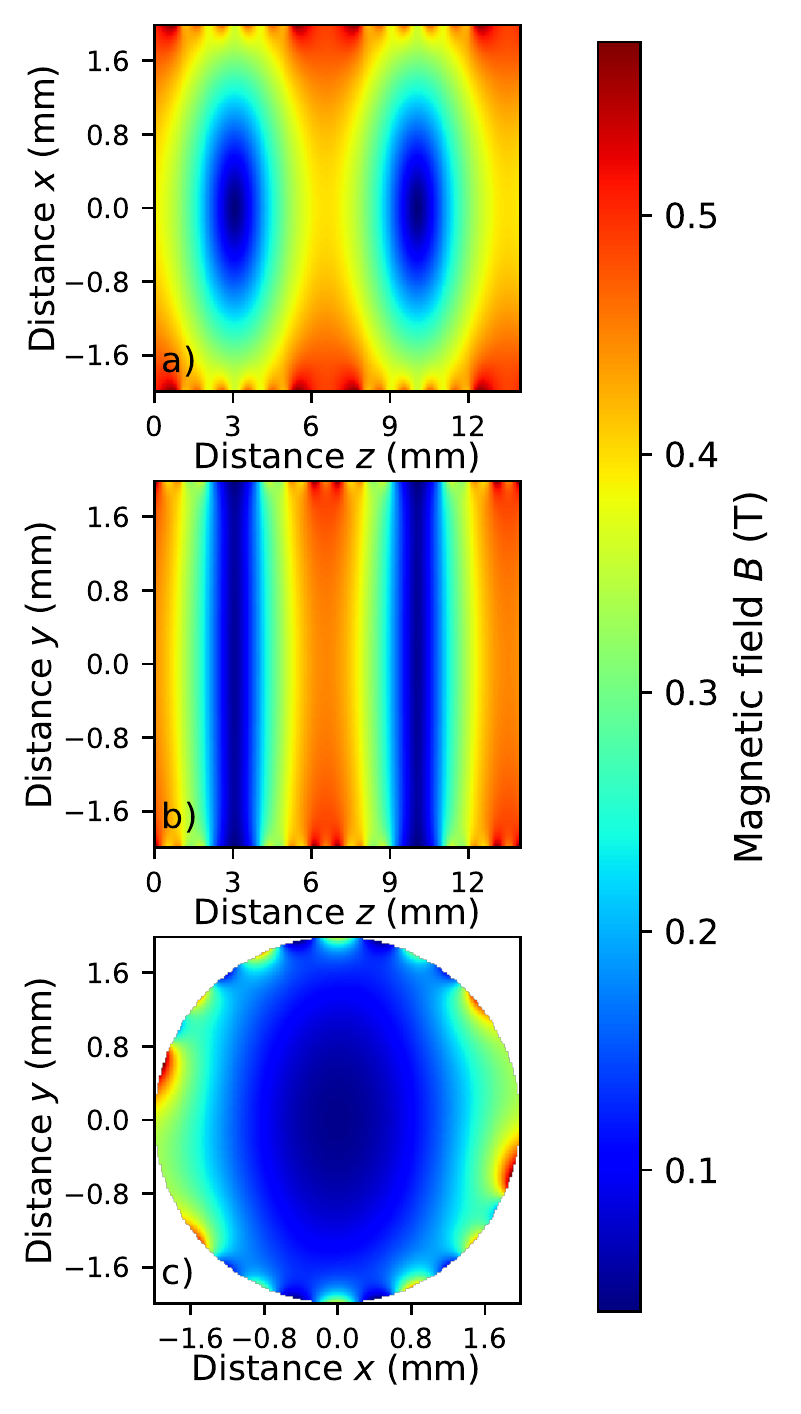}
\caption{Calculated magnitude of the magnetic field a) $|\mathbf{B}(x,0,z,t)|$, b) $|\mathbf{B}(0,y,z,t)|$ and c) $|\mathbf{B}(x,y,0,t)|$ at $t=0$ s generated by currents of the form of $ I_{\mathrm{RH,n}}(t) = I_{\mathrm{RH}}\sin(\phi_{\mathrm{RH}}(t)+n\Delta)$ and $I_{\mathrm{LH,n}}(t) = I_{\mathrm{LH}}\sin(\phi_{\mathrm{LH}}(t)+n\Delta)$ applied to the $n$-th helix. $I_{\mathrm{RH}}=I_{\mathrm{LH}}=+300$~A were assumed as typical values chosen in the experiments.}
\label{fig:rsi_fields_2d}
\end{figure}  

\begin{figure}
\centering
\includegraphics[width=6.6cm]{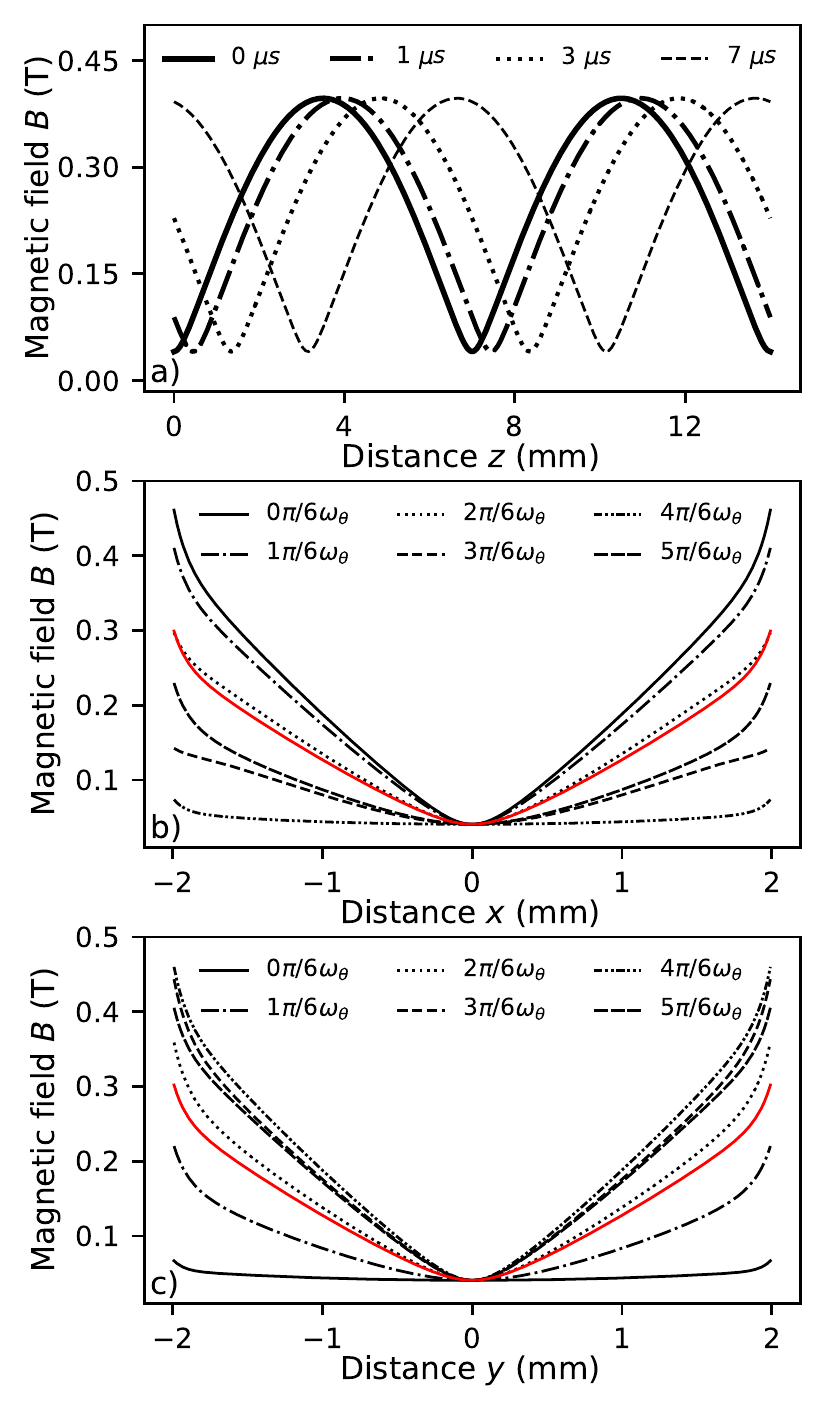}
\caption{Time dependence of the magnetic field. a) Calculated magnetic field $|\mathbf{B}(0,0,z,t)|$ on the central axis of the decelerator at different timestamps (0 $\mu$s, 1 $\mu$s, 3 $\mu$s and 7 $\mu$s) for a travelling wave with velocity 445~m/s. b) and c): Magnetic field in the transverse ($x,y$) directions, b) $|\mathbf{B}(x,0,0,t)|$ and c) $|\mathbf{B}(0,y,0,t)|$, rotating at an angular frequency $\omega_\theta$. The different traces represent the magnetic field at different timestamps over half a period of the rotation. The red trace indicates the magnetic field averaged over a full rotation cycle.  }
\label{fig:rsi_fields_1d}
\end{figure}

\Fref{fig:rsi_fields_2d} a) - c) show the calculated magnetic field $|\mathbf{B}(x,y,z,t)|$ along different directions inside the double-helix Zeeman decelerator generated by the time-dependent currents $I_{\mathrm{RH,n}}(t)$ and $I_{\mathrm{LH,n}}(t)$ over a single helix period $\lambda$. 
The currents assumed are $I_{\mathrm{RH}}=I_{\mathrm{LH}}=+300$~A which correspond to typical values used in the experiments. In a) and b), the magnetic trap with a depth of $\approx$0.5~T is clearly visible along the longitudinal ($z$) direction. In the transverse $x$ direction, the trap depth is $\approx$0.35~T, whereas along the $y$ direction it is only 0.05~T. The travelling-wave aspect of the magnetic field is shown in~\Fref{fig:rsi_fields_1d} a) which shows the calculated magnetic field along the $z$ axis, $|\mathbf{B}(0,0,z,t)|$ at different timestamps. The minima of the magnetic field can be seen to propagate in the $+z$ direction. The corresponding forward velocity of the moving trap is controlled by the phase $\phi_z(t)$ of the currents. With an appropriate choice of parameters $a$ and $b$, the travelling magnetic wave can decelerate, accelerate or propagate with a constant velocity in both the $+z$ and $-z$ directions. The present magnetic-field configuration allows for a 3-dimensional confinement of molecules or atoms in LFS Zeeman states along one of the transverse directions. From~\Fref{fig:rsi_fields_1d} a)-c), it can be seen that at $t$=0~s particles in LFS states can be confined both along the $x$ and $z$ coordinates, while along $y$ the trapping field is more shallow. This limitation can be mitigated by introducing an additional experimentally controllable time-dependent phase $\phi_\theta(t)=\omega_\theta t$. A continuous variation of $\phi_\theta$ leads to a change of the orientation of the radial trapping field, allowing for a dynamic radial trapping of the particles. The effect is illustrated in~\Fref{fig:rsi_fields_1d} b)-c) which display the time-dependence of the magnetic field for both radial directions, $|\mathbf{B}(x,0,0,t)|$ and $|\mathbf{B}(0,y,0,t)|$, at different timestamps over half a period of rotation. The time-averaged field strength over one period, $T=\frac{2\pi}{\omega_\theta}$, is represented by a red trace for both radial components. A detailed characterisation of the dynamics of the travelling trap can be found in Ref. \cite{damjanovic21a}. 

\section{Implementation}
\label{section:Implementation}
\begin{figure}
\centering
\includegraphics[width=\textwidth]{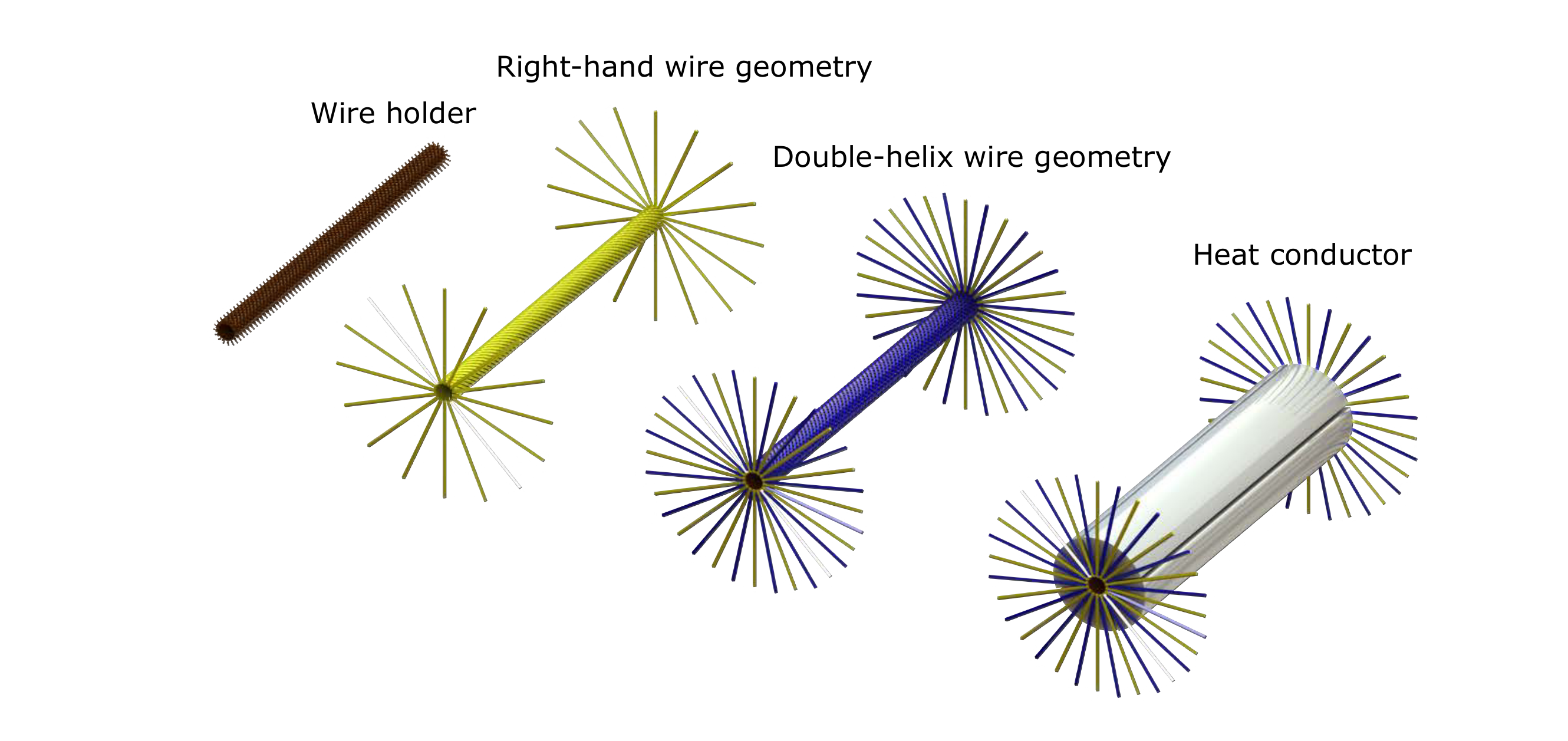}
\caption{Schematic representation of the wire geometry used to generate the traveling magnetic wave. 16 wires (yellow) are wound in right-handed helices around a 4~mm-diameter Vespel tube (brown) superimposed by 16 wires wound in left-handed helices (blue). The coil unit is surrounded by four AlN cases serving as heat conductors.}
\label{fig:wire_scheme}
\end{figure}

The present traveling-wave Zeeman decelerator features a modular design consisting of 16 individual modules. Each module consists of a 56 mm long Vespel tube with a diameter of 4 mm (\Fref{fig:wire_scheme}). Vespel was chosen for its favourable thermal and outgasing properties. On the outer surface of the Vespel tubes, grooves were machined for easier mounting of the wires. On the surface of each tube, a right-handed layer of wires was mounted into the grooves followed by the left-handed layer. Each layer consists of 16 square wires (0.36 mm x 0.36 mm), mounted in a helical geometry with periodicity $\lambda= 14\ $mm. \Fref{fig:wire_scheme} shows a schematic of the Vespel tube (brown) together with a left-handed (yellow) and a right-handed (blue) helical layer of wires. The radii of the right- and left-handed helical layers were chosen to be $R_1 = 2.0$~mm and $R_2=2.4$~mm, respectively.

\begin{figure}
\centering
\includegraphics[width=8.6cm]{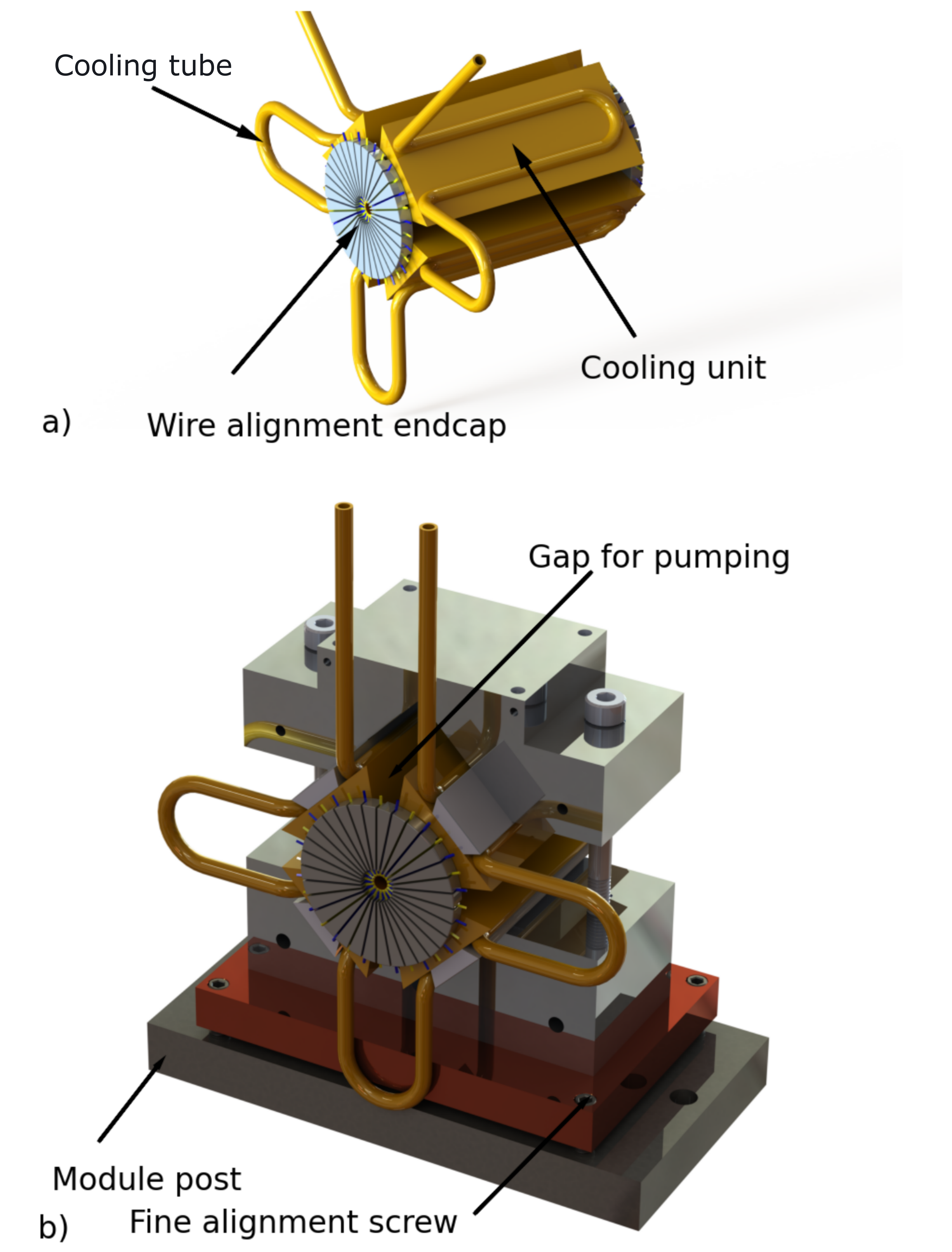}
\caption{Schematic representation of a complete deceleration module. a) Four aluminium-nitride (not shown) and copper cooling units are placed around the wire assembly and cooled by a cooling liquid held at a temperature of 5~\si{\celsius}. The position of each wire on the front and on the back side of each module is fixed by a an alignment endcap. b) Fully assembled deceleration module. The relative alignment of neighbouring modules is adjusted by four alignment screws. The position of each module on a base plate is fixed by alignment pins. }
\label{fig:module_scheme}
\end{figure}

\Fref{fig:module_scheme} shows a schematic of a complete decelerator unit. Four segments of 7~mm thick precision-machined aluminium nitride (AlN) elements were placed around the wires. Aluminium nitride was chosen for its exquisite thermal conductivity of 285~W/mK and its low electric conductance of \num{e-11}-\SI{e-13}{\per\ohm\per\cm}. 
Neighbouring AlN segments were separated by a 0.5 mm gap to enable the evacuation of the different elements of the modules. On top of each AlN segment, a cooling unit machined from 12.5 mm thick copper was placed. The heat produced during the operation of the decelerator was thus dissipated into the AlN and copper cooling units. On the outer surface of the copper unit, a 5 mm wide and 4 mm deep U-shaped groove was machined. Inside the groove, a copper tube with 1.8 mm inner diameter and 4.2 mm outer diameter was placed and connected to the main cooling system. To improve the heat transfer between the cooling units and the tubes, vacuum-compatible heat-conduction paste was applied. An illustration of a fully assembled cooling unit is displayed in~\Fref{fig:module_scheme}(a). All 16 modules of the decelerator were connected in parallel to the main in-vacuum cooling system. The cooling liquid consisting of 50\% deionized water and 50\% propanol and held at a temperature of typically 5~\si{\celsius} was passed through the cooling system by a compact low-temperature thermostat (RCS 6 Lauda, 2.4~kW).

\begin{figure}
\centering
\includegraphics[width=8.6cm,height = 7cm]{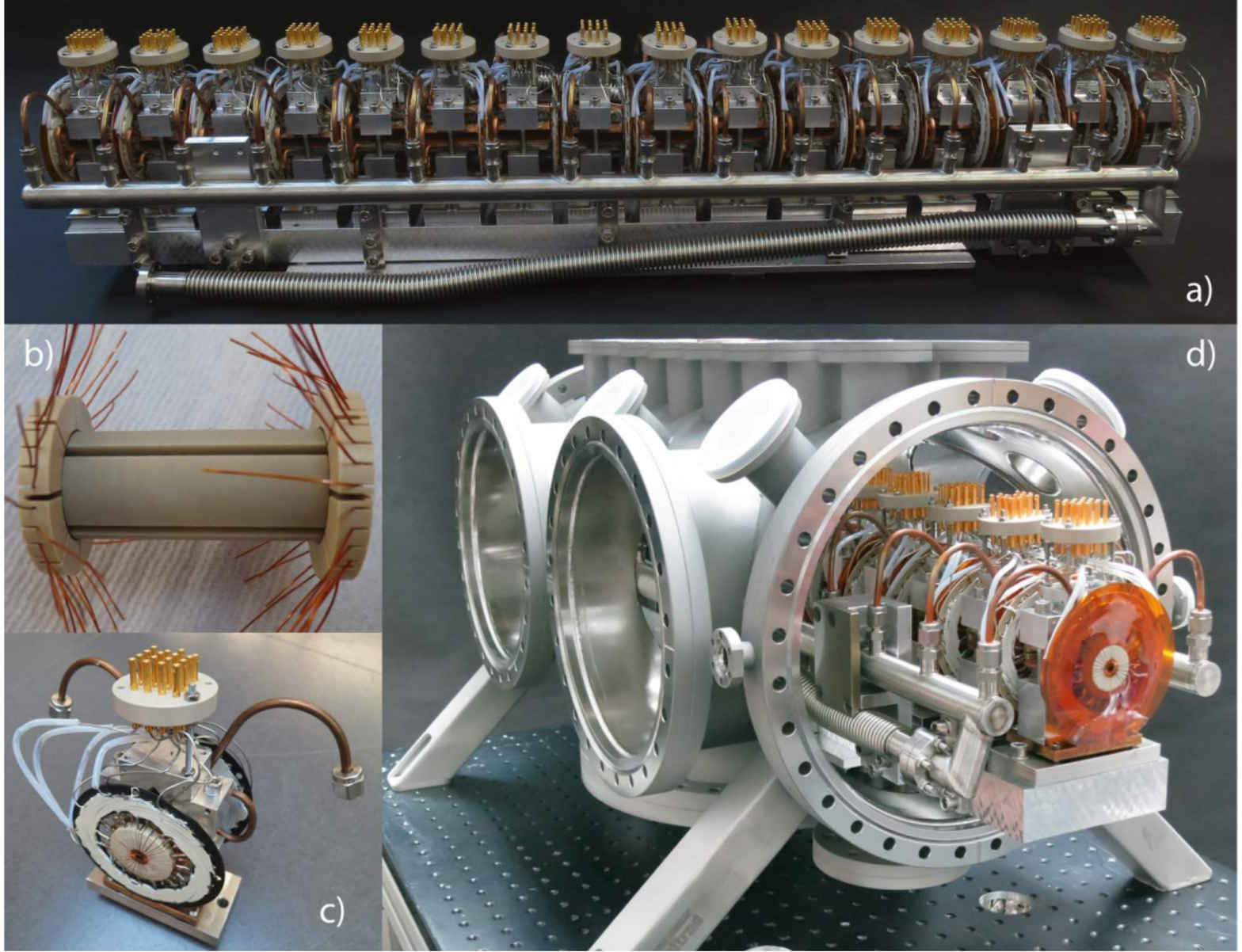}
\caption{Photographs of the decelerator assembly. a) Full decelerator. b) Coil unit. c) Decelerator unit. d) Assembled decelerator in the vacuum chamber.}
\label{fig:decelerator_2}
\end{figure}

On the front and the back side of each module, endcaps enabling a precise alignment of the wires were placed (\Fref{fig:module_scheme} a)). The disk-shaped endcaps were machined from polyether-ether-ketone (PEEK) thermoplastic. 
32 equidistantly spaced slits around each endcap were used to position and guide the individual wires. 
The assembled segments were held by two aluminium braces 
and mounted onto a copper base (60 mm x 43 mm x 6 mm) which was in turn screwed onto an aluminium base (85 mm x 43 mm x 9 mm). Four screws were placed in the four corners of the copper base to enable fine-adjustment of the alignment of the modules, i.e., the vertical offset and angle in the horizontal plane between the neighbouring modules. A fully assembled module is shown in~\Fref{fig:module_scheme} b). Each module was placed on a main stainless steel base plate (100 mm x 896 mm x 30 mm). The position of each module on the base plate was fixed by two alignment pins placed at the bottom of each module (15 mm x 6 mm diameter, 40 mm separation).  \Fref{fig:decelerator_2} a) - d) show photographs of different elements of the decelerator. For the current experiments, a total of 16 decelerator modules was used.

\section{Electronics}
\label{section:Electronics}

The Zeeman deceleration of paramagnetic species requires changes in the magnetic field of several Tesla over distances of a few millimeters within a few microseconds. In addition, these magnetic fields are typically switched at frequencies on the order of tens of kHz. For this purpose, a high-power current generator was developed which produces bipolar currents up to 400 A amplitude with frequencies ranging from 0-40 kHz. 
Bursts of sinusoidal currents $I_\mathrm{L}$ of variable instantaneous frequencies can be generated with a repetition rate $1/t_r$ of typically 2~Hz. 

\begin{figure}
\centering
\includegraphics[width=8.6cm]{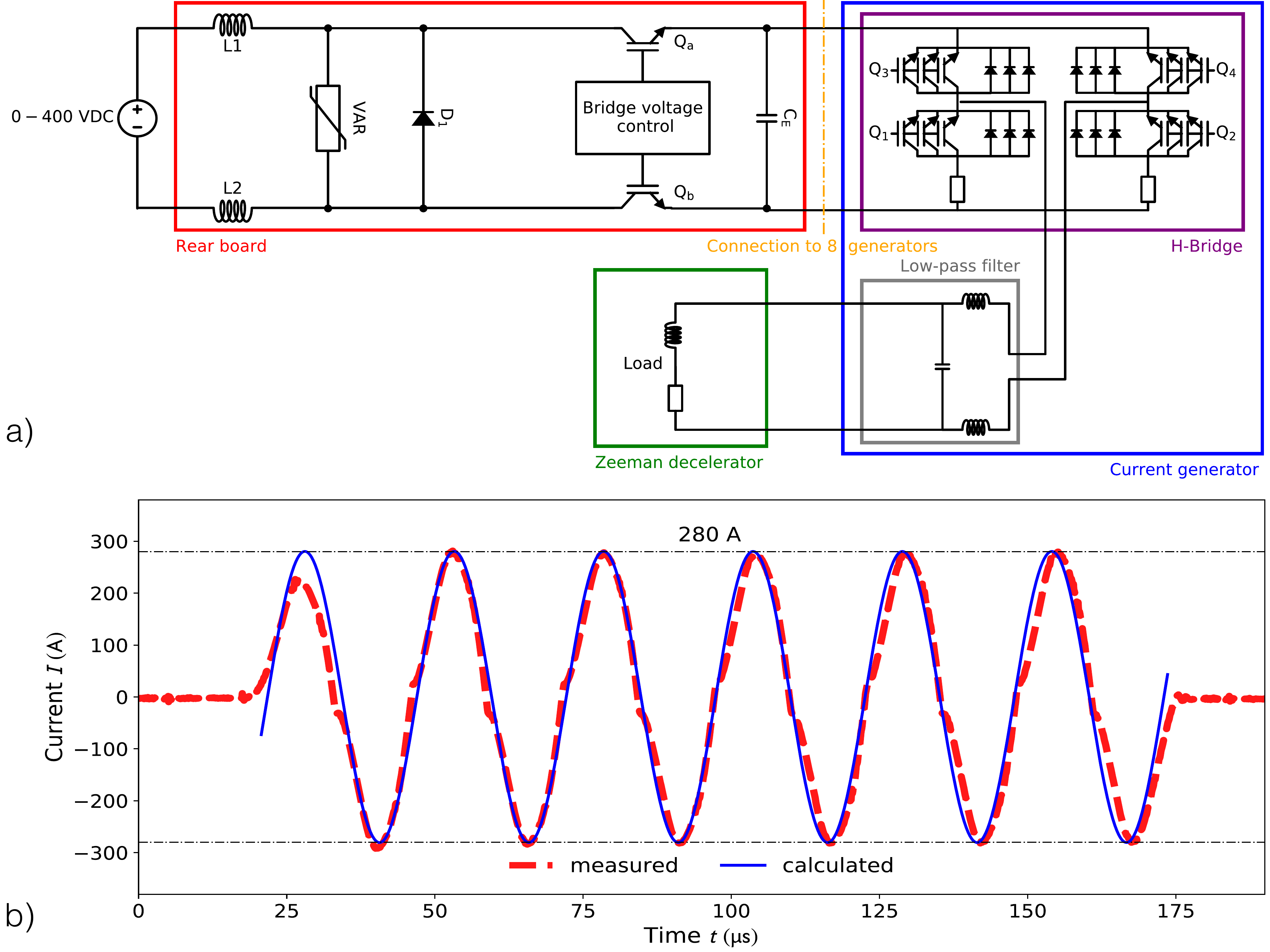}
\caption{a) A simplified schematic diagram of the H-bridge inverter circuit. b) Sinusoidal current pulse realised by operating the generator using a modified square wave control strategy. }
\label{fig:electronics1}
\end{figure}  

The desired current waveform was generated by a full H-bridge inverter \cite{williams02a}, see~\Fref{fig:electronics1} a). 
The H-bridge inverter consists of four switches Q\textsubscript{1}-Q\textsubscript{4} operating the bridge itself. Two additional switches Q\textsubscript{a} and Q\textsubscript{b} are used to control the power supplied to the bridge. Before the very first pulse of each experimental cycle, switches Q\textsubscript{1}-Q\textsubscript{4} are open while switches Q\textsubscript{a}-Q\textsubscript{b} are closed allowing a DC power supply to load the capacitor C\textsubscript{E} (C\textsubscript{E} = 1000 $\mu$F). The DC power supply is smoothly ramped up to avoid current spikes and voltage drops. At the beginning of each pulse, switches Q\textsubscript{a} and Q\textsubscript{b} are opened which isolates the DC power supply from the bridge. This enables to use a single power supply to operate up to eight current generators. Switches Q\textsubscript{1}-Q\textsubscript{4} are then alternately operated in order to convert the DC voltage V\textsubscript{C} of the capacitor C\textsubscript{E} into an AC voltage V\textsubscript{H} across the low-pass filter. A low-pass filter reduces the amplitude of the high-frequency components of V\textsubscript{H} to only a few percent of the main frequency component. After each pulse, switches Q\textsubscript{1}-Q\textsubscript{4} remain open, while Q\textsubscript{a} and Q\textsubscript{b} are closed again, and capacitor C\textsubscript{E} is recharged. 

 \begin{figure}
\vfill
  \begin{center}    
    \includegraphics[width=8.6cm]{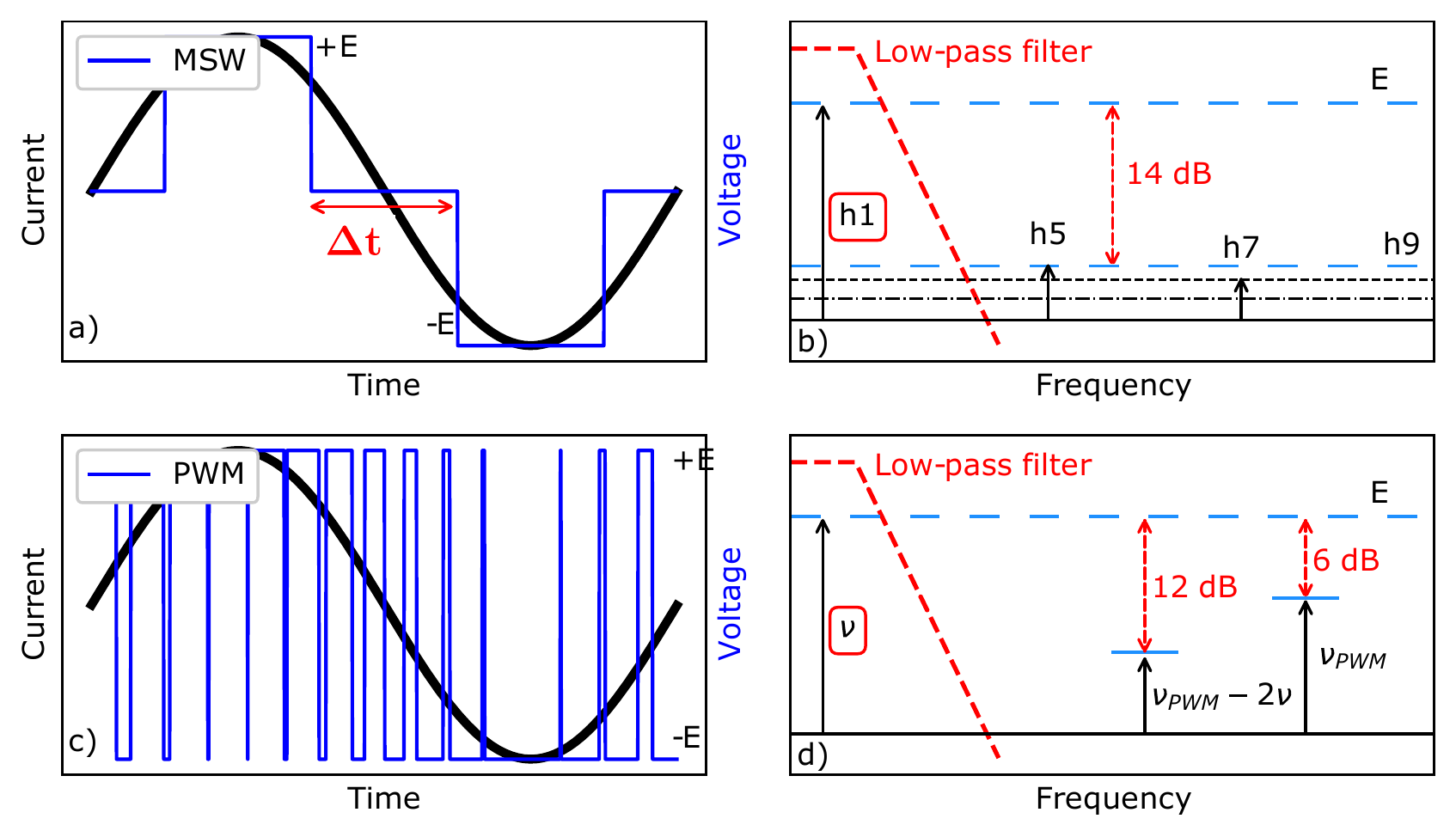}
  \end{center}
  \vfill
\caption{Two strategies for the generation of low- and high-frequency current pulses. a) Three-level modified square wave (MSW): A square pulse consisting of three levels (-E,0,E) at the frequency $\nu$ in the range 20 kHz < $\nu$ < 40 kHz. c) Pulse-width modulation (PWM): Square-wave pulses are generated at a fixed frequency $\nu_{\mathrm{PWM}}$ with modulated duty cycle at a frequency $\nu$ < 20 kHz. In both figures, the voltage dependence across the H-bridge as a function of time is shown as blue traces, and the expected current in the load after filtering is shown as black traces. Schematic power-spectra of the time-dependent voltage $V_H$ across the bridge are shown in b) for MSW and in d) for PWM control. The attenuation function of the low-pass filter is shown schematically by red dashed lines. The amplitudes of higher-frequency components (arrows) are exaggerated for clarity. The low-pass filter is chosen such that any unwanted frequency components are efficiently filtered out.\hspace{\textwidth} \hspace{\textwidth} }
\label{fig:pwm2}
\end{figure}

Two control strategies have been applied to operate the current generator: three-level modified square-wave (MSW) control and pulse-width modulation (PWM), see~\Fref{fig:pwm2}. The MSW control relies on generating a modified square signal of three discrete values (-E,0,+E) at the effective frequency across the load. The ratio between the main frequency component $h1$ and the odd harmonics ($h3, h5, h7, ...$) is tuned via the angle $\beta$ defined as $\beta = \frac{\pi \Delta t}{T}$, where $\Delta t$ is the time duration of the zero level and $T$ is duration of one period. When $\beta = \pi/3$, $h3$ vanishes and the first undesired harmonic becomes $h5$ which is suppressed by 14 dB compared to the fundamental and is easily filtered out. 
In addition, the peak of the current can slightly be tuned via the angle $\beta$ at the expense of an increased harmonic distortion of the signal. MSW control results in only small power losses in the switches, since switching occurs at the frequency $\nu$ of the desired current $I\textsubscript{L}$. This control scheme is especially suitable for the generation of high-frequency currents (20-40 kHz). The harmonics $h5$ and $h7$ lie far outside the bandwidth of the low-pass filter (cutoff at 40 kHz) and are therefore easily eliminated. The generation of low-frequency currents (0-20 kHz) requires a different control strategy because higher harmonics fall into the bandwidth of the low-pass filter. 

The PWM control scheme relies on the application of a square wave at a fixed frequency $\nu_{\mathrm{PWM}}$ to the load, which is chosen much higher than the target frequency $\nu$ of the current $I\textsubscript{L}$. The duty cycle of the square wave is modulated at frequency $\nu$. The voltage $V\textsubscript{H}$ across the bridge then exhibits a fundamental frequency component at $\nu$, and the strongest unwanted frequency components at $\nu_{\mathrm{PWM}}-2\nu$ and $\nu_{\mathrm{PWM}}$. Unwanted frequency components always lie around $\nu_{\mathrm{PWM}}$ and are thus easily pushed outside the bandwidth of the low-pass filter. This control strategy can in principle be adapted to the generation of all frequencies in the range 0-40 kHz, as long as $\nu_{\mathrm{PWM}}$ is chosen high enough. As a disadvantage of this approach, PWM produces more losses in the switches than the MSW method, as more switching is required. As a consequence, we only used the PWM scheme for generating currents with frequencies lower than 20 kHz. 

We have chosen a third-order ($n=3$) Butterworth low-pass filter with frequency response $G(\omega) = 1/\sqrt{1+\omega^{2n}}$, where $\omega$ is the angular frequency. In the MSW control scheme, the filter reduces the harmonics $h5$ by 42 dB, thereby reducing the ratio $h1/h5$ from 20\% to 0.16\%. In the PWM strategy, the minimum acceptable PWM frequency $\nu_{\mathrm{PWM}}$ is set by the frequency component $\nu_{\mathrm{PWM}}-2\nu$ and attenuation requirements. In order to generate frequencies up to 20 kHz, we have set $\nu_{\mathrm{PWM}}$ = 120 kHz. The relative weights of the $\nu_{\mathrm{PWM}}$ and $\nu_{\mathrm{PWM}}-2\nu$ components with the respect to the fundamental are thus 0.4\% and 0.2\%, respectively. 

 The present application requires high peak currents, high switching frequencies and high rates for changing currents $(dI_L/dt>6\ kA/\mu s)$. Therefore, electrical connections needed to be treated with extra care. The capacitor $\mathrm{C_E}$ is mounted on a motherboard with a biplanar connection to the H-bridge. In order to minimize the voltage overshoots due to the connection inductance, the biplanar connection is realized through low-impedance printed-circuit boards which exhibit a linear induction lower than 1 nH/m. For the same reason, the low pass filter is located as close as possible to the load and consists of only three components: two inductors and one capacitor. Due to the high currents flowing in the low-pass-filter, the internal magnetic field in these conductors is quite high, so low inductances of 1.2 $\mathrm{\mu H}$ and air-core inductors were preferred in order to minimize core losses. Finally, the connection from the H-bridge to the load is about 1 m long and consists of a 50 $\mathrm{\Omega}$ coaxial cable. This cable introduces a small additional inductance of 250 nH/m which has been taken into account in the design of the low-pass filter. 
 
 The electronics are controlled by a LabView code \cite{bitter2006a} developed in-house. From the initial velocity of the molecular beam and desired final velocity of the molecular package at the end of the decelerator, the parameters $a$, $b$ and $c$ are calculated which define the phase $\phi_z(t)$ at time $t$ controlling the positions of the traveling traps (see~\Fref{section:A traveling magnetic wave}). These parameters together with the predefined rotational frequency of the traveling trap $\omega_{\theta} = 2\pi f$ ($f$ = 0-10 kHz) are then sent to the current generators. An embedded software for the current generators accepts the parameters and uses the parameters to calculate the current switching sequences for each generator. 
 Additionally, extensive status-checking functionalities, e.g. for monitoring the temperature of the boards, the status of the switching sequences and possible overvoltage and overcurrent failures, are in place for each generator to verify its correct operation.

~\\~\\
\section{Experiment}
\label{section:experiment}

\begin{figure*}[!t]
\centering
\vspace{20pt}
\includegraphics[width=15.4cm, height=10.5cm]{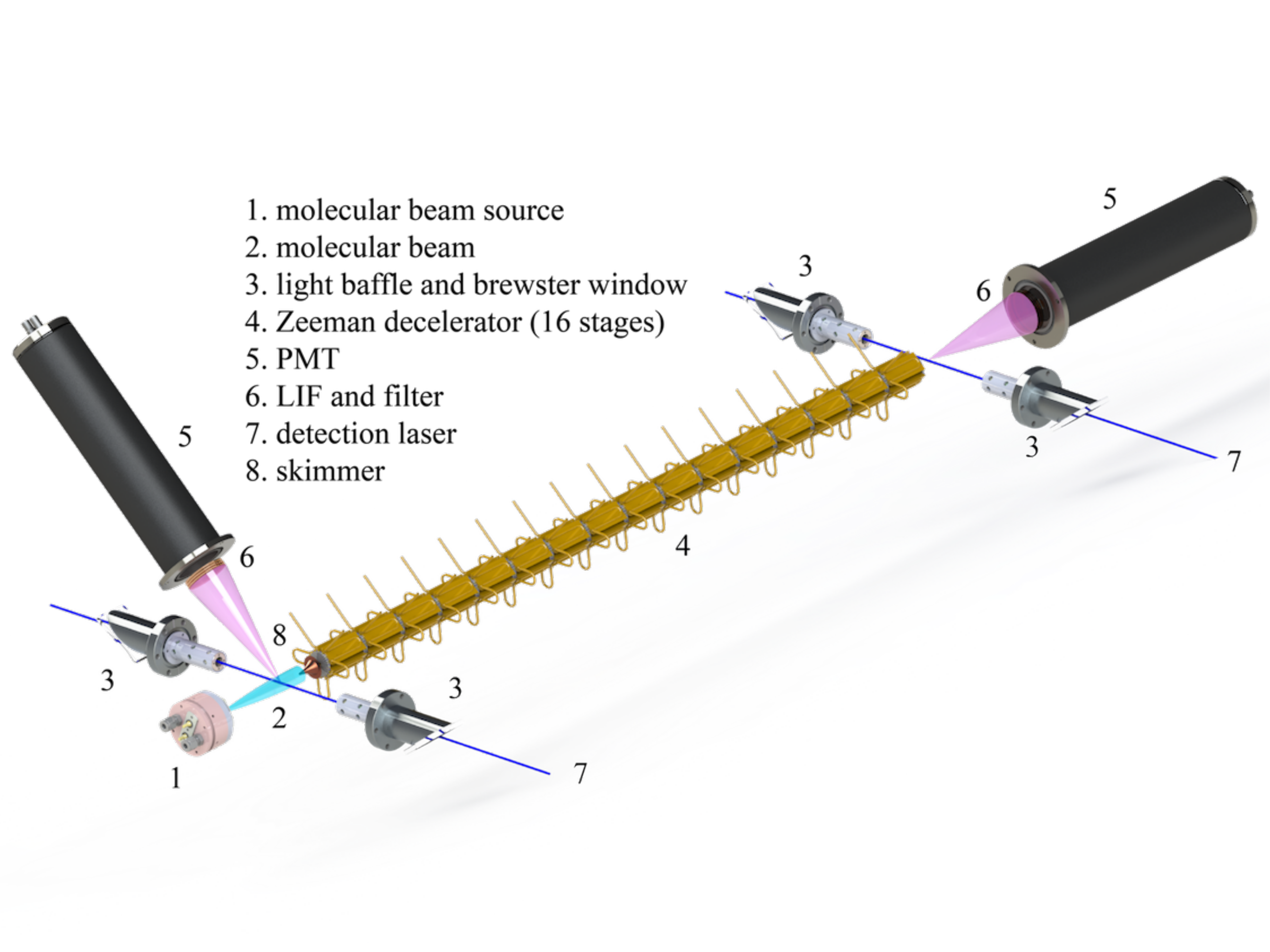}
\caption{Schematic of the present experimental apparatus. A molecular beam of OH radicals is produced in an electric discharge of H$_2$O in Xe from a pulsed gas nozzle (1). After passing through a skimmer (2), the beam is coupled into the traveling-wave Zeeman decelerator (3), where it is decelerated from an initial velocity of 445~m/s down to variable finial velocities. Following deceleration, the molecules are detected by laser-induced fluorescence (LIF) collected onto a photo-multiplier tube (PMT)(6).}
\label{fig:decelerator}
\end{figure*}  

A schematic of the complete experimental apparatus is shown in~\Fref{fig:decelerator}. A molecular beam of OH radicals with a mean longitudinal velocity of 445~m/s, and 50~m/s velocity spread (FWHM) was produced by a pulsed gas valve discharging H\textsubscript{2}O seeded in Xe \cite{ploenes16a}. In the supersonic expansion emanating from the valve, about 98\% of the molecules were populated in the lowest rotational level $(J=3/2)$ of the vibrational $(v=0)$ and electronic ground state $(X^2\Pi_{3/2})$~\cite{ploenes16a}. In the presence of a magnetic field, the degeneracy between different $m_J$ levels $(m_J = \pm3/2,\pm1/2)$ is lifted, where $m_J$ denotes the magnetic quantum number. Only particles in the  LFS states $(m_J=+1/2,+3/2)$ are amenable to Zeeman deceleration~\cite{vanhaecke07a}. The effective magnetic moment of OH in the $m_J=+3/2$ state is $ 1.4\mu_B$ \cite{maeda2015a}, where $\mu_B$ denotes the Bohr magneton. The molecular beam passed through a skimmer (Beam Dynamics, 2~mm diameter) 7~cm downstream from the valve orifice to enter the deceleration chamber where it was coupled into the decelerator.

After deceleration, the OH molecules in the $X^2\Pi_{3/2}, v=0, J=3/2$ state were excited by a 282~nm  laser pulse (with the energy of 1.2~mJ) into the $A^2\Sigma^{+},v=1, J=1/2$ excited state. The laser radiation was generated by the output of a frequency-doubled dye laser operating with a Rhodamin 6G dye pumped by a Nd:YAG laser. The resulting fluorescence from the $A^2\Sigma^{+}(v=1)\rightarrow X^2\Pi(v=1)$ transition (720~ns lifetime) centered at 313~nm was collected by a UV-grade lens with 50~mm focal length and detected by a photo-multiplier tube (Electron Tubes B2/RFI, 9813 QB). Unwanted light was filtered by a combination of four UV-grade filters~\cite{scharfenberg12}. 
Time-of-flight (TOF) profiles of the molecules exiting the decelerator were recorded by monitoring the LIF yield as a function of the laser delay. 
All TOFs were referenced to the opening of the valve. The time step between consecutively recorded TOF data points was 2 $\mu s$ and each point was averaged over 200 laser shots. 

\section{Results and discussion}

\begin{figure}[htpb]
\centering
\includegraphics[width=0.5\textwidth, height=10cm]{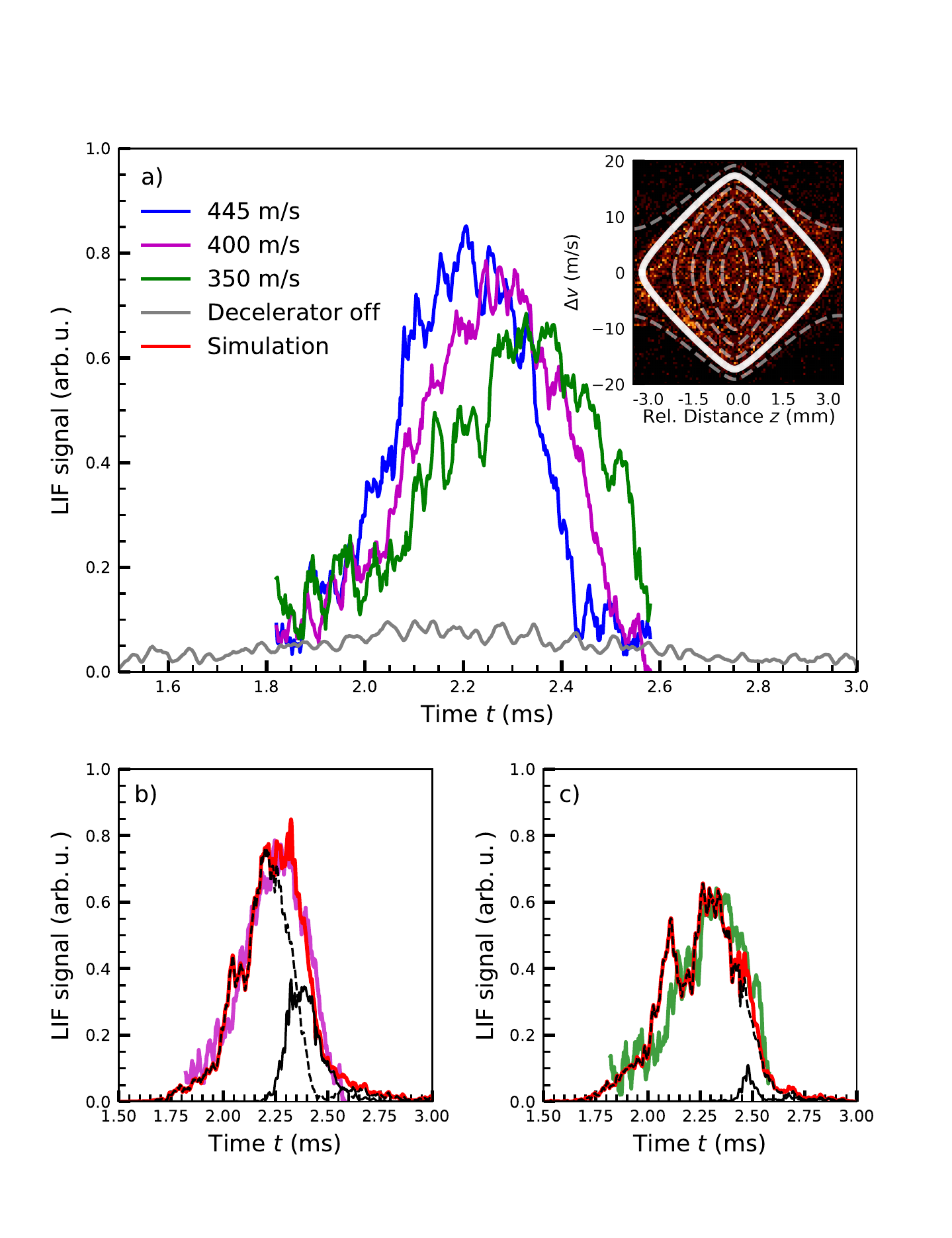}
\caption{a) Time-of-flight (TOF) measurements of decelerated OH radicals obtained by collecting laser-induced fluorescence (LIF ) 36~mm downstream from the decelerator. The molecules entered the decelerator with an initial velocity of 445 m/s and were decelerated to three different final velocities: 445 m/s (guiding mode, blue), 400 m/s (magenta), 350 m/s (green). The grey trace represents a measurement when the decelerator was switched off. Inset: 1D phase-space distribution of particles along the deceleration axis inside a single traveling trap. The separatrix is indicated as a solid white line and isoenergetic trajectories in phase space separated by $\approx 0.05$ cm$^{-1}$ are represented by dashed white lines. b),c) Comparison of experimental TOF traces (magenta trace: final velocity 400 m/s; green trace: final velocity 350 m/s) to TOF traces extracted from numerical simulations (red traces). The black traces show the contributions to the overall TOF from particles that are efficiently confined inside traveling traps during the deceleration process (full line) and those which are not (dashed line).}
\label{fig:TOF_main}
\end{figure}

TOF profiles of OH molecules decelerated to specific final velocities starting from an initial velocity of 445~m/s are shown in~\Fref{fig:TOF_main} a). 
The blue trace corresponds to a guiding experiment in which the final velocity of the traveling wave is matched to the initial velocity so that the molecules were guided at 445~m/s around the node of the traveling magnetic field. The magenta and green traces correspond to experiments with a constant deceleration of the magnetic trap from the initial velocity down to 400~m/s and 350~m/s, respectively. The grey curve represents the TOF profile obtained when the decelerator was switched off. The inset in~\Fref{fig:TOF_main}a) shows a 1D phase-space distribution (PSD) along the deceleration axis of particles which were efficiently trapped inside a single traveling trap throughout the deceleration process extracted from numerical trajectory simulations  \cite{damjanovic21a}. In the numerical trajectory simulations, $3.6\times10^5$ molecules were created 9.3 cm upstream from the decelerator. The initial positions and velocities of the molecules were sampled from a normal distribution. Molecules were created with 445 m/s mean forward velocity and 50 m/s longitudinal velocity spread (FWHM) and a longitudinal spatial spread of 60 mm. The molecules were propagated to the detection region with their positions and velocities recorded every 100 microseconds. The inset of~\Fref{fig:TOF_main}a) shows a snapshot taken 2.2~ms after the molecules exit the nozzle at time they arrive in the detection region. The thick white curve represents the separatrix, a line encircling the area of PSD for which molecules are efficiently trapped during deceleration, corresponding to the trap depth of 0.24 cm$^{-1}$ (0.34 K). Broken white lines in the inset indicate isoenergetic phase-space trajectories separated by 0.05 cm$^{-1}$. One can see that particles were confined inside a $\approx$7~mm long region along the decelerator axis which corresponds to the geometric size of the trap defined by a $\lambda$/2 =7 mm. 

In Ref.~\cite{trimeche11a}, it was shown that it is possible to discern contributions from particles that are effectively trapped around the node of the magnetic wave from those that accidentally distribute around it in the time-of-flight profiles. In our case, this was not possible due to experimental limitations. In order to obtain an isolated structure of decelerated molecules in the time-of-flight measurements, a molecular beam with a narrow spatial spread along the deceleration axis would be required so that only a single traveling trap would be occupied with molecules. In the present experiments, the entirety of the molecular beam was overlapped with travelling traps and thereby multiple traps were filled with molecules. This explains the broad TOF profile of the decelerated beam shown in~\Fref{fig:TOF_main} a). 

These limitations could be circumvented by, e.g., employing laser dissociation of HNO\textsubscript{3} for generating a spatially narrow beam of OH radicals\cite{vandemeerakker05a} instead of an electric discharge. 

However, it is also possible to characterize the decelerated fraction of the molecular beam by comparison with simulations. \Fref{fig:TOF_main} b) and c) illustrate such a comparison of  experimentally obtained TOF profiles to those extracted from numerical particle-trajectory simulations. The red traces correspond to numerical simulations for final velocites of b) 400~m/s and c) 350~m/s. 
To understand how decelerated molecules contribute to the overall TOF profiles, the position and velocity of each molecule in the simulation were checked if they fall within phase-space accepted area of each trap. The total simualted TOF profile is then split into two parts arising from decelerated and undecelerated molecules shown as full and broken black lines in~\Fref{fig:TOF_main}, respectively. An evolutionary algorithm was developed for optimizing the initial parameters of the molecular beam in order to minimize the discrepancy between numerical simulations and experimental results. Details can be found in Ref. \cite{damjanovic21a}. The overall agreement between experiments and simulations are good. Small discrepancies are attributed to the lack of a detailed knowledge of the properties of the molecular beam, notably its exact spatial and temporal profile before it is coupled into the decelerator. 

\begin{figure}[htpb]
\centering
\includegraphics[width=8.5cm, height=10cm]{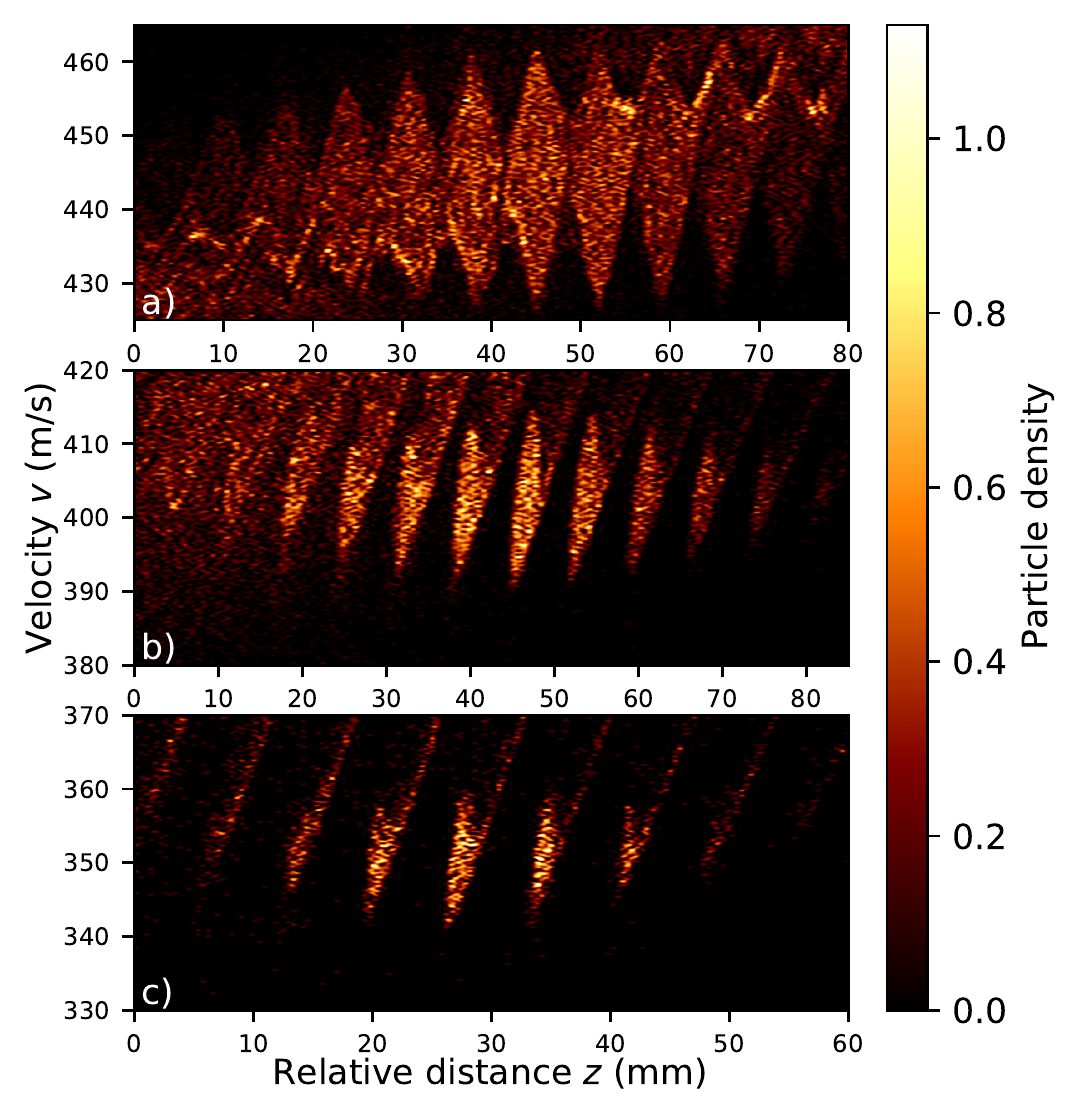}
\caption{ 1D phase-space distribution of particles along the deceleration axis extracted from numerical trajectory simulations at the point in time at which particles are exiting the decelerator and entering into the detection region. The molecules were decelerated from initial velocity of 445~m/s down to a) 445~m/s, b) 400~m/s  and c) 350~m/s. The normalized particle density is represented by a color-map. }
\label{fig:Phase_space}
\end{figure}    

\Fref{fig:Phase_space} presents PSDs of particles extracted from the numerical trajectory simulations shown in~\Fref{fig:TOF_main}. Only molecules which are effectively trapped during flight are shown. The normalized particle density is represented by a color map. Panel a) correspond to the PSD of particles guided at 445~m/s extracted at a timestamp of 2.2 ms in the simualtion when they exit the decelerator and enter the detection region. Panels b) and c) represent PSDs of particles decelerated to final velocities of 400~m/s and 350~m/s and extracted at 2.3 ms and 2.4 ms, respectively. Each decelerator module consists of 8 geometrical traps, but depending on when currents are switched on (off) relative to the arrival (departure) of the package of molecules, differing numbers of effective traps can be created leading to a different fractions of the molecular beam being coupled into the decelerator. This is reflected in the present PSDs. In guiding mode (\Fref{fig:Phase_space} a)), 10 traps filled with molecules are propagated through the decelerator. On the other hand, lower final velocities lead to lower effective trap depths which in turn lead to smaller fractions of molecules being decelerated (b,c).

\section{Conclusions and outlook}
In conclusion, we have designed, implemented and characterized a new type of travelling-wave Zeeman decelerator. 
We have demonstrated the successful operation of the apparatus by decelerating a molecular beam of OH molecules in the $X^2\Pi_{3/2}$ state from an initial forward velocity of 450~m/s down to a final velocity of 350~m/s. Experimental time-of-flight traces were favorably compared with numerical trajectory simulations.

The decelerator is modular in design allowing for its ready extension. Compared to conventional Zeeman or Stark decelerators \cite{scharfenberg09a,dulitz15a,wiederkehr10a,bethlem00a}, the present implementation exhibits a full three-dimensional confinement of the molecules as also evidenced from the trajectory simulations leading to an improved overall phase-space acceptance \cite{damjanovic21a}. For the purpose of producing the required time-dependent currents, we have developed compact arbitrary waveform current generators capable of producing currents up to 300~A and 0-40~kHz frequency. The decelerator was so far operated at 2~Hz repetition rate. With improvements to the cooling system, the repetition rates could be increased up to 10~Hz. 

In the future, the decelerator will be extended by another 16 modules, allowing for improved deceleration capabilities and the deceleration of OH radicals down to final velocities of about 50~m/s. Even at the present length of the decelerator with 16 modules, light particles such as H atoms and metastable H$_2^*$ can be decelerated to effectively a standstill. We expect that the present decelerator will be ideally suited for the investigation of scattering experiments between cold molecules and trapped ions or loading decelerated molecules into a trap. 
\\

\section*{Acknowledgments}
We thank the mechanical and electronics workshops from FHI and the Department of Chemistry at the University of Basel for their support. Funding from the Swiss National Science Foundation, grant nr. 200020\_175533, and the University of Basel is acknowledged. D.Z. acknowledges the financial support from Freiwillige Akademische Gesellschaft (FAG) Basel, the Research Fund for Junior Researchers of the University of Basel and the National Key R\&D Program of China (No. 2019YFA0307701). 




\section*{References}
\bibliographystyle{iopart-num}
\bibliography{ref} 
\end{document}